\documentclass[runningheads]{llncs}
\usepackage[T1]{fontenc}
\usepackage{multirow}
\usepackage{graphicx}
\usepackage[pagebackref=true,breaklinks=true,colorlinks,bookmarks=false]{hyperref}
\begin{document}
\title{Abdominal multi-organ segmentation in CT using
Swinunter }
%
%
\author{Mingjin Chen\inst{1}\ \and
Yongkang He\inst{1} \and
Yongyi Lu\inst{1}
} 
\authorrunning{First Author Name here et al.}
%
\institute{Guangdong University of Technology, Guangdong 51006, China \and
\\
\email{\{2112103033\}@mail2.gdut.edu.cn}}
\maketitle              
\begin{abstract}
Abdominal multi-organ segmentation in computed tomography (CT) is crucial for many clinical applications including disease
detection and treatment planning. Deep learning methods have shown
unprecedented performance in this perspective. However, it is still quite challenging to accurately segment different organs utilizing a single network due to the vague boundaries of organs, the complex background, and the substantially different organ size scales. In this work we used make transformer-based model for training. It was found through previous years' competitions that basically all of the top 5 methods used CNN-based methods, which is likely due to the lack of data volume that prevents transformer-based methods from taking full advantage. The thousands of samples in this competition may enable the transformer-based model to have more excellent results. The results on the public validation set also show that the transformer-based model can achieve an acceptable result and inference time.

\end{abstract}

\section{Introduction}
Automated medical image segmentation techniques ~\cite{huo20193d} have
shown prominence for providing an accurate and reproducible solution for organ segmentation. Recently, deep learning-based human organs segmentation techniques ~\cite{myronenko20193d,jiang2020two,myronenko2020robust,isensee2021nnu} have achieved state-of-the-art performance in various benchmarks ~\cite{simpson2019large,bakas2018identifying}. These advances are mainly due to the powerful feature extraction
capabilities of Convolutional Neural Networks (CNN)s. However, the limited kernel size of CNN-based techniques restricts their capability of learning long-range
dependencies that are critical for accurate segmentation of tumors that appear
in various shapes and sizes.

Although several efforts ~\cite{chen20193d} have tried to address
this limitation by increasing the receptive field of the convolutional kernels, the
effective receptive field is still limited to local regions.

In order to extract more effective local and global contextual representations, Swin UNETR~\cite{hatamizadeh2021swin} have been proposed as a hierarchical vision transformer that computes self-attention in an efficient shifted window partitioning scheme. Swin UNETR  utilizes a U-shaped network with
a Swin transformer as the encoder and connects it to a CNN-based decoder at different resolutions via skip connections.

In this report we used a transformer-based model for training. It was found through previous years' competitions that basically all of the top 5 methods used CNN-based methods, which is likely due to the lack of data volume that prevents transformer-based methods from taking full advantage. The thousands of samples in this competition may enable the transformer-based model to have more excellent results.

\section{Method}

\subsection{Preprocessing}
We first crop the non-zero regions of the image and resample
the cropped data, and then we use Z-Score standardization to normalize the data. The Z-Score standardized formula is as follows:

\begin{equation}
    z=\frac{x-\mu}{\sigma} 
\end{equation}

where $\mu$ is the average value of the CT value of the image label, and $\sigma$ is the variance of the CT value of the image label.


\subsection{Network}
\begin{figure}[htbp]
\centering
\includegraphics[scale=0.20]{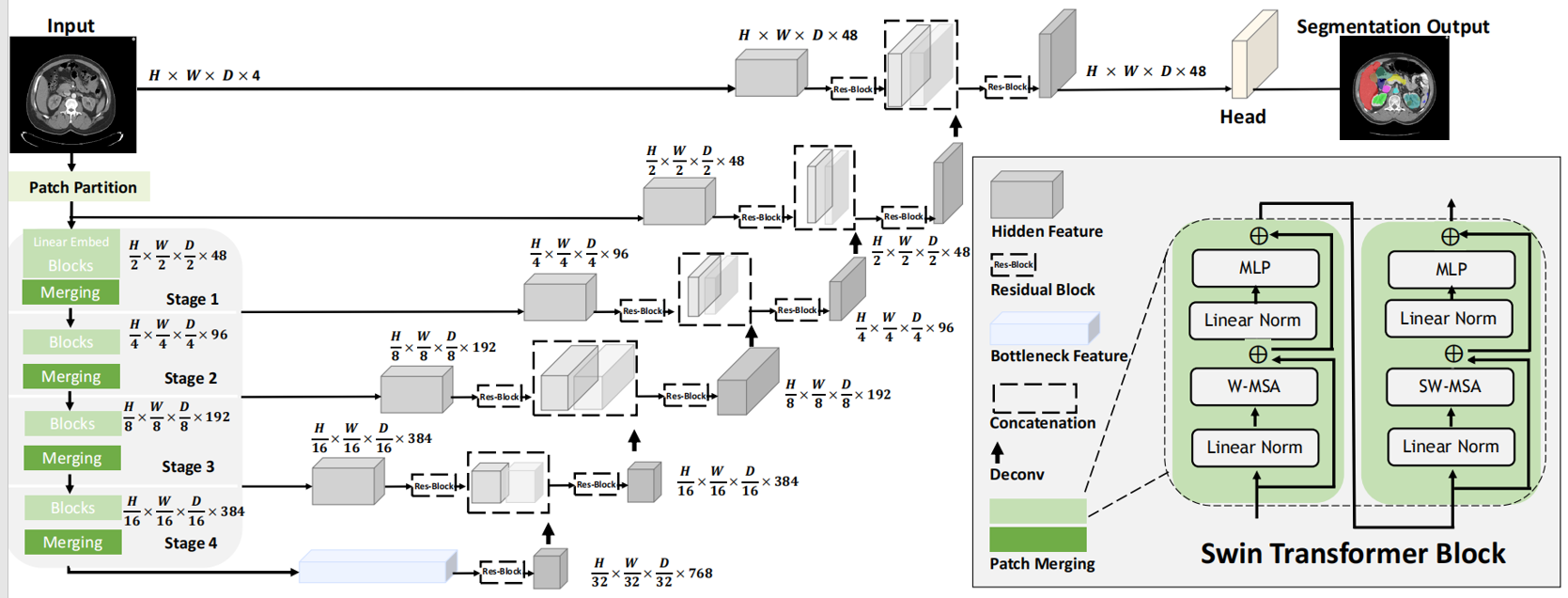}
\caption{Overview of the Swin UNETR architecture.
}
\label{fig:Network}
\end{figure}

We use \textbf{Swin UNTER}~\cite{hatamizadeh2021swin} as our network for training and testing. It is consist of a 
Transformer encoder and a CNN decoder. As shown in Fig.\ref{fig:Network}, the input to the Swin UNETR model
is a token with a patch resolution of $(H, W, D)$.

\subsubsection{Encoder}
Swin UNTER has a U-shaped network design in which the extracted feature representations of the encoder are used in the decoder via skip connections at each
resolution. The self-attention is computed into non-overlapping windows that are created in the partitioning stage for efficient token interaction modeling. It  utilize windows of size
$M\times M \times M$ to evenly partition a 3D token into $[\frac{H'}{M}]\times [\frac{W'}{M}] \times [\frac{D'}{M}]$ regions at a given layer $l$ in the transformer encoder. Subsequently, in layer $l + 1$, the partitioned window regions are shifted by $[\frac{M}{2}]\times [\frac{M}{2}] \times [\frac{M}{2}]$ voxels. In subsequent layers, the regular and window partitioning multi-head
self-attention (MSA) modules are used to learn effective local and global contextual representations.

\subsubsection{Decoder}
At each stage $i (i \in {0, 1, 2, 3, 4})$ in the encoder and the bottleneck $(i = 5)$, the output feature representations are reshaped into size $\frac{H}{2^i}]\times \frac{W}{2^i} \times \frac{D}{2^i}$ and fed into a residual block comprising of two $3\times3\times3$ convolutional layers that are normalized by instance normalization~\cite{ulyanov2016instance} layers. Subsequently, the resolution of the feature maps are increased by a factor of 2 using a deconvolutional layer and the outputs are concatenated with the outputs of the previous stage.
The concatenated features are then fed into another residual block as previously
described.  The final segmentation outputs are computed by using a $1 \times 1 \times 1$
convolutional layer and a sigmoid activation function.

\subsection{Loss function}
we use the soft Dice loss functions, it has been proven to be robust in various medical image segmentation tasks~\cite{LossOdyssey}. It is computed in a voxel-wise manner as 

\begin{equation}
    L(G, Y)=1-\frac{2}{J}\sum_{j=1}^{J}\frac{\sum_{i=1}^{I} G_{i,j}Y_{i,j}}{\sum_{i=1}^{I}G^2_{i,j}+\sum_{i=1}^{I}Y^2_{i,j}}   
\end{equation}
where $I$ denotes voxels numbers; $J$ is classes number; $Y_{i,j}$ and $G_{i,j}$ denote the probability of output and one-hot encoded ground truth for class $j$ at voxel $i$, respectively.

\subsection{Post-processing}
In some computer vision tasks, it is necessary to do some
post-processing on the output of the model to optimize the visual effect, and connected domain is a common post-processing method. Especially for segmentation tasks, sometimes there are some false positives in the output mask. Finding independent contours with small area through 3D connected domain and
removing them can effectively improve the visual effect. We use connected domain principal component analysis to remove 3D small connected domains and
retain the largest part of each label connected domain.

\section{Experiments}
\subsection{Dataset and evaluation measures}
The FLARE 2023 challenge is an extension of the FLARE 2021-2022~\cite{MedIA-FLARE21}\cite{FLARE22}, aiming to aim to promote the development of foundation models in abdominal disease analysis. The segmentation targets cover 13 organs and various abdominal lesions. The training dataset is curated from more than 30 medical centers under the license permission, including TCIA~\cite{TCIA}, LiTS~\cite{LiTS}, MSD~\cite{simpson2019MSD}, KiTS~\cite{KiTS,KiTSDataset}, and AbdomenCT-1K~\cite{AbdomenCT-1K}. The training set includes 4000 abdomen CT scans where 2200 CT scans with partial labels and 1800 CT scans without labels. The validation and testing sets include 100 and 400 CT scans, respectively, which cover various abdominal cancer types, such as liver cancer, kidney cancer, pancreas cancer, colon cancer, gastric cancer, and so on. The organ annotation process used ITK-SNAP~\cite{ITKSNAP}, nnU-Net~\cite{nnUNet}, and MedSAM~\cite{MedSAM}.

The evaluation metrics encompass two accuracy measures—Dice Similarity Coefficient (DSC) and Normalized Surface Dice (NSD)—alongside two efficiency measures—running time and area under the GPU memory-time curve. These metrics collectively contribute to the ranking computation. Furthermore, the running time and GPU memory consumption are considered within tolerances of 15 seconds and 4 GB, respectively.

\subsection{Implementation details}
\subsubsection{Environment settings}
The development environments and requirements are presented in Table~\ref{table:env}.

\begin{table}[!htbp]
\caption{Development environments and requirements.}\label{table:env}
\centering
\renewcommand\arraystretch{1.6}{
\begin{tabular}{ll}
\hline
System       & Ubuntu 18.04.5 LTS or Windows 11\\
\hline
CPU   & AMD EPYC 7502 32-Core Processor \\
\hline
RAM                         &16$\times $4GB; 2.67MT$/$s\\
\hline
GPU (number and type)                         & Four NVIDIA V100 16G\\
\hline
CUDA version                  & 11.0\\                          \hline
Programming language                 & Python 3.7.12\\ 
\hline
Deep learning framework &  torch 1.12.0, torchvision 0.13.0\\
\hline
Specific dependencies         &                       monai 0.9.0  \\                                                                      
\hline
Code     &   \url{https://github.com/chanwendy/FLARE23}                                                             \\
\hline

\end{tabular}
}
\end{table}

\subsubsection{Training protocols}
The Training protocols are presented in Table 2. First, we use only partially labeled images to train a pretrain model, then we use this pretrain model to generate pseudo-labels for the unlabeled images as our labels, and then we use this pseudo-labeled as well as partially labeled data to continue to train the pretrain model to obtain our final model. For data enhancement we first convert all the images to the same spacing and then apply rotation, translation, and sampling of positive samples. For the training of the model we used AdamW as an optimizer with a weight decay of 0.05

\begin{table*}[!htbp]
\caption{Training protocols.}
\label{table:training}
\begin{center}
\renewcommand\arraystretch{1.4}{
\begin{tabular}{ll} 
\hline
Network initialization         & “he" normal initialization \\
\hline
Batch size                    & 2 \\
\hline 
Patch size & 96$\times$96$\times$96  \\ 
\hline
Total epochs & 1000 \\
\hline
Optimizer          &   AdamW     \\ \hline
Initial learning rate (lr)  &  1e-4 \\ \hline
Lr decay schedule & None \\
\hline
Training time                                           & 72.5 hours \\  \hline 
Loss function & Dice \\     \hline
Number of model parameters    & 244.47M \\ \hline
Number of flops & 59.32G \\ \hline
CO$_2$eq & 1 Kg \\  \hline
\end{tabular}
}
\end{center}
\end{table*}


\section{Results and discussion}
The average running time is 10s per case in inference phase. The maximum used GPU memory is 14129MB. Table 3 lists the results on the validation set. Overall, better results are achieved for larger and regular organs like the liver and the kidney. Worse results are achieved for smaller and complex organs like the gallbladder and the duodenum. These results indicate that it is difficult to handle the size variations utilizing the Swinunter, and specific modules should be designed to particularly address the issue.

\begin{figure}[!htbp]
\centering
\includegraphics[scale=0.50]{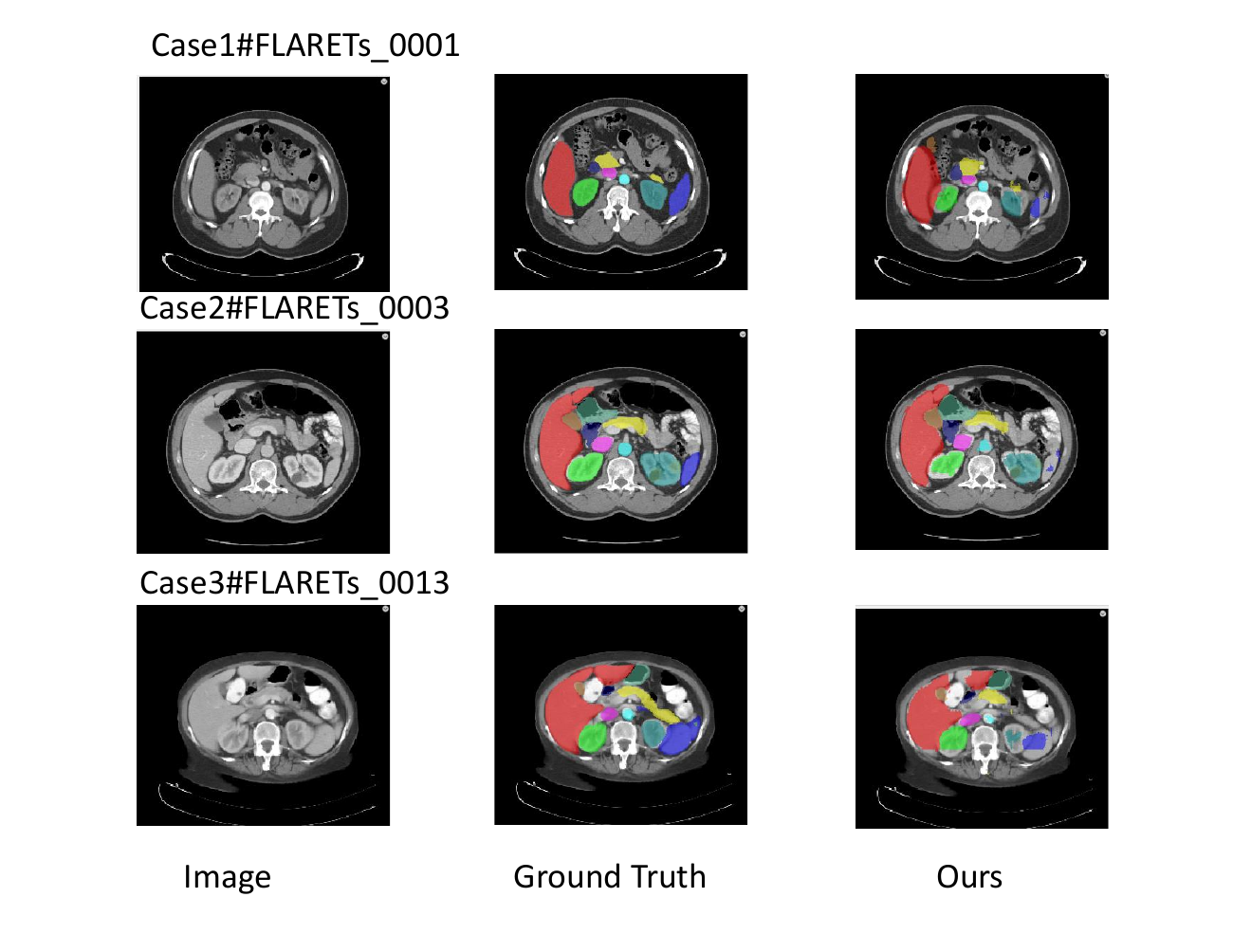}
\caption{ Example segmentation results with high segmentation accuracy.
}
\label{fig:seg}
\end{figure}

\begin{table}[htbp]
\caption{The dice metrics of the prediction results of Swinunter on the open source validation set and The dice metrics of their embedding prediction results on the validation set. 
}
\centering
\renewcommand\arraystretch{1.2}{
\begin{tabular}{l|cc|cc|cc}
\hline
\multirow{2}{*}{Target} & \multicolumn{2}{c|}{Public Validation} & \multicolumn{2}{c|}{Online Validation} & \multicolumn{2}{c}{Testing} \\ \cline{2-7} 
                        & DSC(\%)            & NSD(\%)           & DSC(\%)            & NSD(\%)           & DSC(\%)      & NSD (\%)     \\ \hline
Liver                   & 0.8969 $\pm$ $0.098$                  &   0.6019 $\pm$ 0.078                  &   -                 &     -                &            &              \\
Right Kidney            &    0.8800 $\pm$ 0.196                &   0.6045 $\pm$ 0.078                &      -              &    -               &              &              \\
Spleen                  &    0.8741 $\pm$ 0.147                &       0.6013$\pm$ 0.077            &          -          &    -               &              &              \\
Pancreas                &    0.8271 $\pm$ 0.091                &     0.6021 $\pm$ 0.077              &     -               &  -                 &              &              \\
Aorta                   &   0.877 $\pm$ 0.121                 &     0.6049 $\pm$ 0.077              &      -              &   -                &              &              \\
Inferior vena cava      &   0.8620 $\pm$ 0.115                 &        0.6077 $\pm$ 0.077           &       -             &   -                &              &              \\
Right adrenal gland     &      0.7252$\pm$0.232              &     0.6079$\pm$0.078              &          -         &    -               &              &              \\
Left adrenal gland      &    0.7192$\pm$0.223                &       0.6075 $\pm$0.078            &          -          &    -               &              &              \\
Gallbladder             &    0.7954$\pm$0.253                &     0.6078 $\pm$0.078              &           -         &      -             &              &              \\
Esophagus               &    0.7650$\pm$0.167                &        0.6067$\pm$0.078           &          -          &     -              &              &              \\
Stomach                 &  0.9051$\pm$0.072                  &      0.6066$\pm$0.078             &           -         &        -           &              &              \\
Duodenum                &   0.8360$\pm$0.093                 &       0.6074$\pm$0.078            &             -       &       -            &              &              \\
Left kidney             &    0.8820$\pm$0.157                &         0.6075$\pm$0.078          &               -     &      -             &              &              \\
Tumor                   &    0.7642$\pm$0.186                &      0.6067$\pm$0.078             &           -         &      -             &              &              \\ \hline
Average                   &  0.8293$\pm$0.061                  &        0.6058$\pm$0.002           &         -           &       -            &              &              \\ \hline
\end{tabular}
}
\end{table}
\subsection{Quantitative results on validation set}
We use swinunter as our main network framework to first train our pretrain model with only partially labeled data, and then predict labels without labeled data as pseudo-labels by pretrain model. Finally the final model is obtained by fine-tuning the pretrain model with a mixture of pseudo-labeled data and partially labeled data. The final results on the publicly available 50 validation sets are shown in Table 3.

\subsection{Qualitative results on validation set}
Fig. 2 presents the segmentation results of three cases for which satisfactory segmentation accuracy is achieved. Fig. 3 shows the segmentation results of three cases for which low segmentation accuracy is achieved. It can be observed that for those easy samples, the background is quite simple, whereas for hard samples, the background is quite complex. Meanwhile, the most obvious problem for the low accuracy is the under-segmentation of the small organs and the over-segmentation of the big organs. 
Possible reasons could be that the model
training is not sufficient or that the model complexity is not enough due to the
small training data we utilized. I believe more accurate results can be obtained
if we exploit the unlabeled data in an effective way.

\begin{figure}[!htbp]
\centering
\includegraphics[scale=0.5]{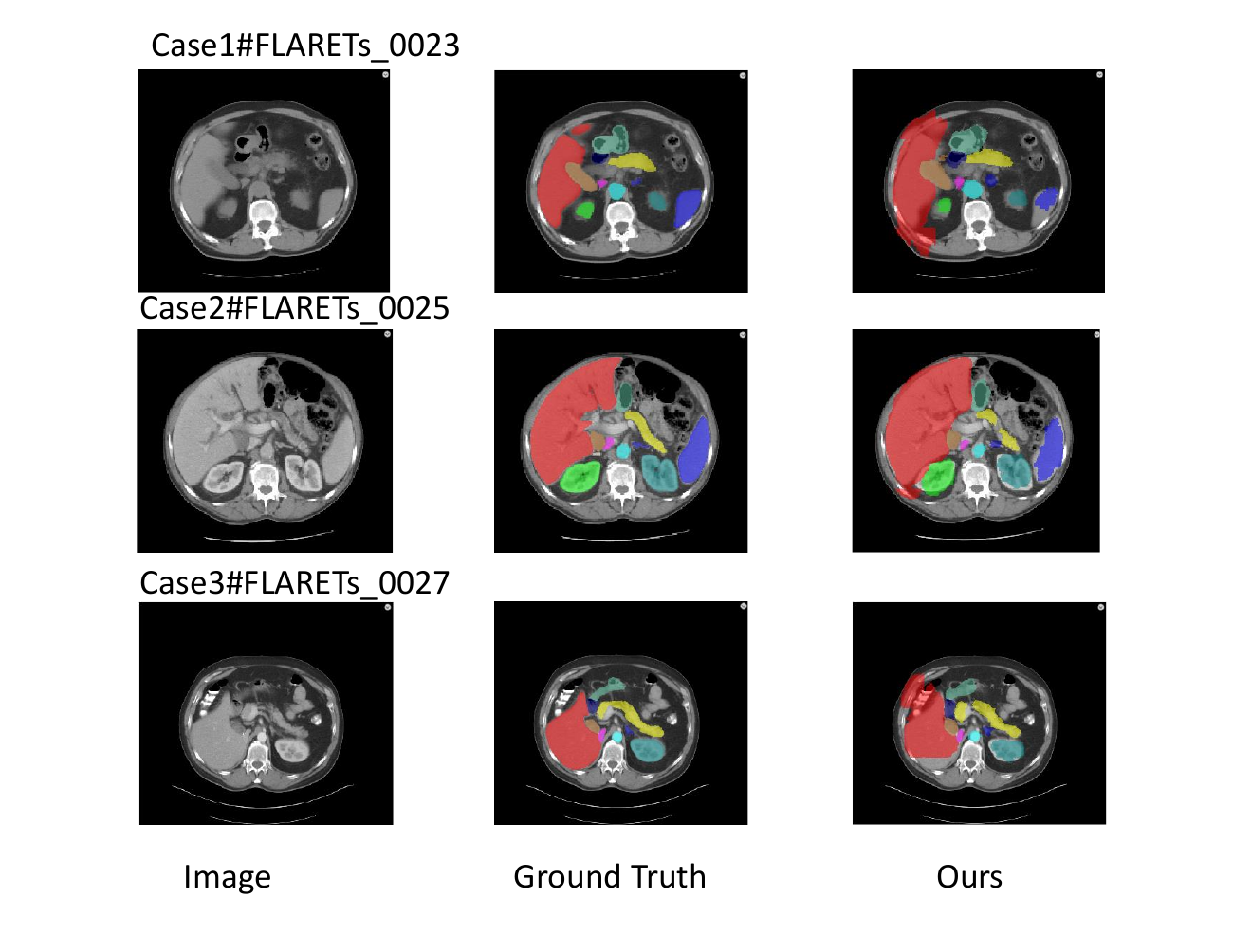}
\caption{ Example segmentation results with low segmentation accuracy.
}
\label{fig:segbad}
\end{figure}

\subsection{Limitation and future work}
Although we have thousands of training samples, the fact that the labels are partially labeled and half of the data is pseudo-labeled leads to unstable training when fine-tuning the pretrain model, which can easily cause the model to overfit into one of the organs. 

In future work, we will make a partial decoder for partially labeled data to decode a specific organ, and for the problem of pseudo-labeling accuracy, we will consider some newer methods to update the pseudo-labeling accuracy to get a better fine-tuned model.

\section{Conclusion}
We used a transformer-based model as our main framework because the amount of data for this competition was large enough, compared to last year, so we thought that a large amount of data might be more suitable for a transformer-based model. The results of our model have an acceptable effect on the open source validation set as well as the speed of inference. Since our current work does not require excessive accuracy of pseudo-labeling, which is our drawback, we need to further consider the accuracy of pseudo-labeling and the correct decoding of some of the labels in our future work.

\subsubsection{Acknowledgements} The authors of this paper declare that the segmentation method they implemented for participation in the FLARE 2023 challenge has not used any pre-trained models nor additional datasets other than those provided by the organizers. The proposed solution is fully automatic without any manual intervention. We thank all the data owners for making the CT scans publicly available and CodaLab~\cite{codalab} for hosting the challenge platform.

%
%
%
\bibliographystyle{splncs04}
\bibliography{ref}

\newpage
\begin{table}[!htbp]
\caption{Checklist Table. Please fill out this checklist table in the answer column.}
\centering
\begin{tabular}{ll}
\hline
Requirements                                                                                                                    & Answer        \\ \hline
A meaningful title                                                                                                              & Yes/No        \\ \hline
The number of authors ($\leq$6)                                                                                                             & Number        \\ \hline
Author affiliations, Email, and ORCID                                                                                           & Yes/No        \\ \hline
Corresponding author is marked                                                                                                  & Yes/No        \\ \hline
Validation scores are presented in the abstract                                                                                 & Yes/No        \\ \hline
\begin{tabular}[c]{@{}l@{}}Introduction includes at least three parts: \\ background, related work, and motivation\end{tabular} & Yes/No        \\ \hline
A pipeline/network figure is provided                                                                                           & Figure number \\ \hline
Pre-processing                                                                                                                  & Page number   \\ \hline
Strategies to use the partial label                                                                                             & Page number   \\ \hline
Strategies to use the unlabeled images.                                                                                         & Page number   \\ \hline
Strategies to improve model inference                                                                                           & Page number   \\ \hline
Post-processing                                                                                                                 & Page number   \\ \hline
Dataset and evaluation metric section is presented                                                                              & Page number   \\ \hline
Environment setting table is provided                                                                                           & Table number  \\ \hline
Training protocol table is provided                                                                                             & Table number  \\ \hline
Ablation study                                                                                                                  & Page number   \\ \hline
Visualized segmentaiton example is provided                                                                                     & Figure number \\ \hline
Limitation and future work are presented                                                                                        & Yes/No        \\ \hline
Reference format is consistent.  & Yes/No        \\ \hline

\end{tabular}
\end{table}

\end{document}